# TensorFlow with user friendly Graphical Framework for object detection API

Heemoon Yoon[1], Sang-Hee Lee[1], Mira Park[1]

[1]Discipline of Information, Communication and Technology, School of Technology, Environments and Design, University of Tasmania, Hobart TAS7001, Australia

Heemoon Yoon (heemoon.yoon@utas.edu.au), Sang-Hee Lee (sanghee.lee@utas.edu.au),
Mira Park (mira.park@utas.edu.au)

**Abstract**

TensorFlow is an open-source framework for deep learning dataflow and contains application programming interfaces (APIs) of voice analysis, natural language process, and computer vision. Especially, TensorFlow object detection API in computer vision field has been widely applied to technologies of agriculture, engineering, and medicine but barriers to entry of the framework usage is still high through command-line interface (CLI) and code for amateurs and beginners of information technology (IT) field. Therefore, this is aim to develop an user friendly Graphical Framework for object detection API on TensorFlow which is called TensorFlow Graphical Framework (TF-GraF). The TF-GraF provides independent virtual environments according to user accounts in server-side, additionally, execution of data preprocessing, training, and evaluation without CLI in client-side. Furthermore, hyperparameter setting, real-time observation of training process, object visualization of test images, and metrics evaluations of test data can also be operated via TF-GraF. Especially, TF-GraF supports flexible model selection of SSD, Faster-RCNN, RFCN, and Mask-RCNN including convolutional neural networks (inceptions and ResNets) through GUI environment. Consequently, TF-GraF allows anyone, even without any previous knowledge of deep learning frameworks, to design, train and deploy machine intelligence models without coding. Since TF-GraF takes care of setting and configuration, it allows anyone to use deep learning technology for their project without spending time to install complex software and environment.

## 1. Introduction

Deep learning is a part of machine learning, its architectures such as deep neural network (DNN), convolutional neural networks (CNN), and others have been applied to computer vision, speech recognition, natural language processing [1]. Especially, the computer vision has been rapidly developed accompany with the deep learning technology which is applied in various industry fields of engineering [2], agriculture [3], and medicine [4]. The computer vision technologies based on the deep learning for classification and detection in images are largely classified as classification, object detection, and segmentation, of these, the object detection is defined as detection of multiple objects in a single image [5]. However, it is difficult to develop the deep learning model for the object detection without understanding of computer science knowledge because it comes with an extensive collection of tools, platforms, and software but none of them provides user friendly graphical user interface (GUI). Additionally, developers should be required expert programming skills to use deep learning technologies since building a deep learning model involves quite intensive programming. Accordingly, increasing of object detection technologies demand, the deep learning frameworks such as TensorFlow [6], Keras [7], Pytorch [8], and Caffe [9] for object detection have been developed by various companies.

One of the most popular deep learning frameworks in these days is TensorFlow which is the end to end opensource machine learning platform [10]. It provides useful libraries and resources for building and evaluation models [10]. Furthermore, it offers high-performance application programming interfaces (APIs) to build DNN and CNN based models such as voice analysis, natural language process, and computer vision [10]. However, as non-information technology (IT) people, there are many difficult aspects to access the frameworks. It starts with an installation which is the most basic steps for using any kinds of software. There are numerous ways of TensorFlow installation and the easiest and reliable one is to install it on a virtual machine, but this virtual machine is an unfamiliar environment for non-IT people. Since this installation process holds various possible ways to get errors such as version mismatch or operating system mismatch, the installation process needs to be done in very delicate manner. On the top of this, extra dependencies such as OpenCV [11], Numpy [12], and Cython [13] and environmental programs such as Python or Anaconda are needed to be installed within same environment.

After the installation, there are tremendous steps for building a model such as adding the users' data in the relevant folder which can be recognised by TensorFlow for generating TensorFlow readable input files, setting paths of the output files, and configuring those files for every execution. These steps should be done by typing of long commands in use the command line interface (CLI). Using CLI is also very challengeable task for non-IT people but command-line arguments are the primary mechanism for communicating with TensorFlow.

TensorFlow also provides very useful multi-model evaluation metrics but the users should manage various file types (e.g., ckpt, pb, and config) for using those functions. Unless user has IT background, or even some people with IT background, this process can be quite challenging.

This hardship makes people unable to utilize deep learning technology even before start. Also, even if the process had been done successfully, installation process requires afterward maintenance works of the system which means user need to have another extra knowledge to do so. Alternatively, there are some of visual programming software like Rapid Miner [14], Orange [15], Dataikiu [16], Knime [17], and AIDeveloper [18]. The visual programming software, however, require users to spend time to learn the software and they are not intuitive.

Therefore, we are aiming to develop a deep learning tool called TensorFlow with user friendly Graphical Framework for object detection API (TF-GraF) that exposes the entire TensorFlow object detection API through an intuitive GUI. TensorFlow is one of numerous open-source frameworks for CNN [6], and it is a free library from Google for computing operations on tensors which is very popular for building neural networks and machine learning and this framework is the main platform of TF-GraF. TensorFlow is very powerful API for the deep learning and TF-GraF makes these functions available through an intuitive GUI for many researchers who cannot manage programming-based interfaces since it does not require any knowledge of programming. If the labelled data is available, models can be trained, tested, and analyzed directly in the TF-GraF.

## 2. Implementation of TF-GraF

*2.1. GUI implementation*

TF-GraF was implemented to replace command line interface to graphical user interface to provide user friendly interface for using TensorFlow API. In order to improve the interface, typing and executing commands which were originally done by Secure Shell (SSH) client program such as Putty were replaced to clicking buttons within GUI views. To help understanding of users, TF-GraF divided its functions into step by step flows and presented with intuitive icons. In addition, TF-GraF was designed to be able to check the outcome of actions of users visually after they finish each step. The GUI components used for TF-GraF was implemented using Java Swing component.

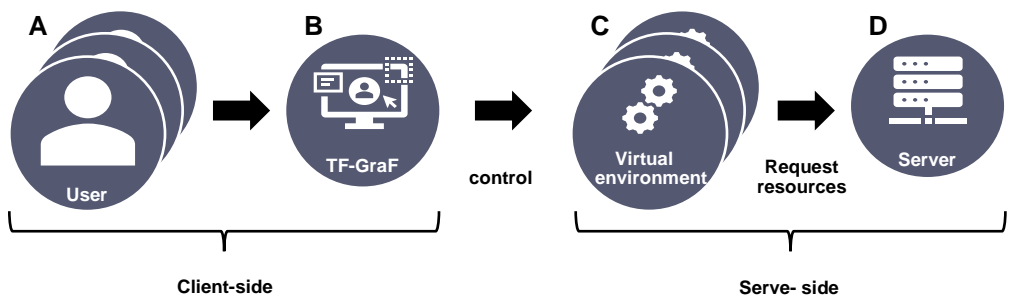

**Figure 1.** System overview of TensorFlow with user friendly Graphical Framework for object detection API (TF-GraF). Access of multiple user access (A) to TF-GraF in client-side (B), created virtual environments in server-side (C), requesting of resources to server by virtual environment (D)

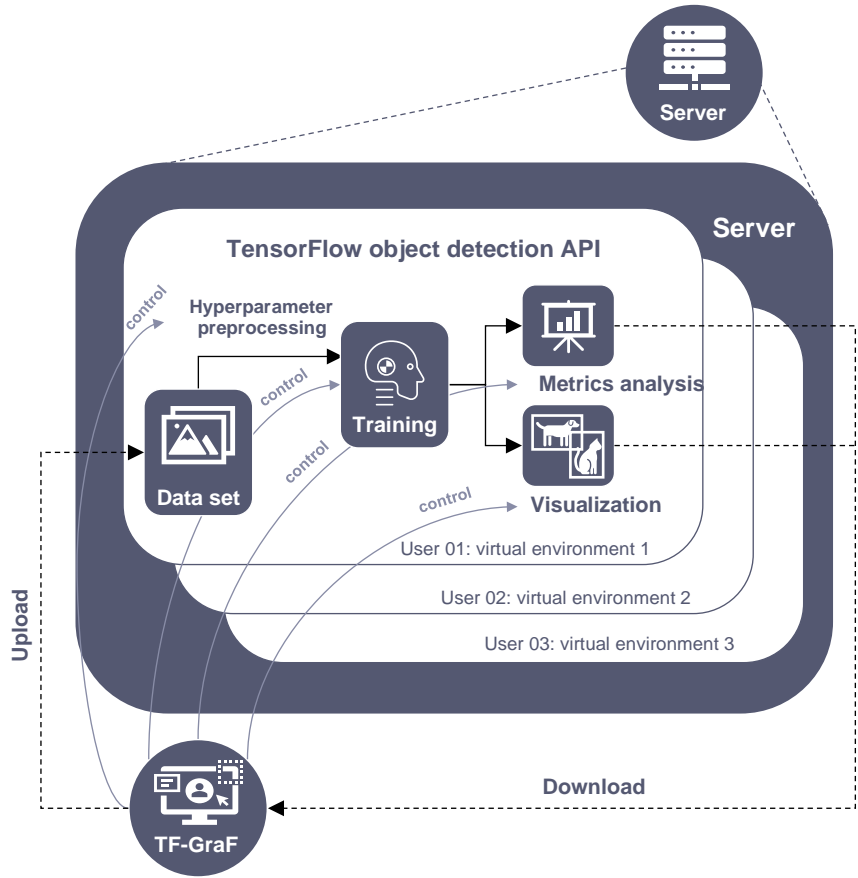

**Figure 2.** Server-side overview of TensorFlow with user friendly Graphical Framework for object detection API (TF-GraF). Each virtual environment contains independent TensorFlow object detection API and each user has separated virtual environment. The architectures (SSD, Faster-RCNN, RFCN, and Mask-RCNN) and backbones (MobileNets, Inceptions, ResNets, and Inception-ResNets) were implanted in server-side which can be selected in client-side through TF-GraF. Uploading of annotated image data and downloading of trained models, results of metrics analysis, and visualization of test images by trained models are controlled by TF-GraF.

*2.2. Implementation overview*

The TF-GraF consists of two main parts (Figure 1) which are client-side (Figure 1A and 1B) and server-side (Figure 1C and 1D). The function of client-side is mainly triggering commands in the server-side and returning result of server-side output to display that in designated views. Users can visually control the GUI to access and manipulate their account in this Client-side (Figure 1A and 1B).

On the other side, server-side implementation holds TensorFlow object detection API environment. Server-side provides operations of TensorFlow object detection API functions and allows user to avoid from complicated installation process. This is possible because server-side provides fully preinstalled virtual environments of TensorFlow object detection API for new users (Figure 1C). Server-side takes most functions of deep learning operation as server generally has better capacity of computing power. In summary, server-side provides deep learning operations and client-side provides communication and displays.

*2.3. Server-side implementation*

As shown in Figure 2, server-side consists of several layers of virtual environment. First, server itself must have a certain level of hardware and system for TensorFlow object detection API implementation. Inside of the server, virtual environments are independently implemented for each user. In this virtual environments, users can upload their data such as images and annotated files to the server for training models and they also can download trained model, the results of metric analysis and object detected images by trained models to their local device via TF-GraF. Virtual environments are the place where dependencies for TensorFlow environments

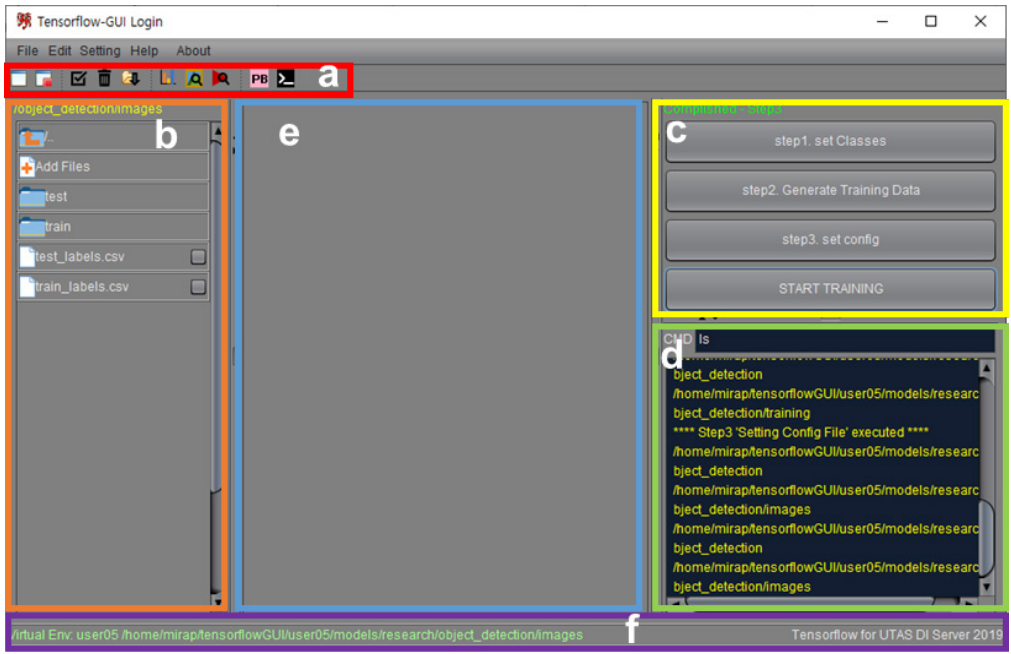

**Figure 3.** Overview of TensorFlow with user friendly Graphical Framework for object detection API (TF-GraF). **a**, Toolbar view, **b**, display current directory and files view, **c**, control training related steps view, **d**, control server with command line, **e**, display selected image view, and **f**, display current directory path and activated environment view, creating of labelmap files

are installed. Each user has their assigned virtual environment and that can be accessed by entering password so that uploaded data can be kept secure and privacy Since each user has their own virtual environment, users can operate their TensorFlow without disturbing or being disturbed by other users' setting. Since each user has their own virtual environment, users can operate their TensorFlow without disturbing or being disturbed by other users' setting. Even though it looks like waste of memory storage in server, maintaining separated virtual environments is more beneficial in terms of maintainability because this allows administrators to handle issues without affecting other users' environments. In addition to this, customizing their own virtual environment up to their research subject is another benefit which can be achieved from this since each virtual environment has individual TensorFlow settings which can be customized up to users request to administrator.

Finally, inside of the virtual environment, TensorFlow API is installed. This environment layer is where TensorFlow object detection API operates. When user trigger command by clicking buttons on GUI from client-side, this layer will be triggered to operate designated function. In this layer, most of TensorFlow object detection API functions such as selection of architectures (SSD, Faster R-CNN, RFCN, and Mask-RCNN), backbones (MobileNets, Inceptions, ResNets, and Inception-ResNet), data augmentation, hyperparameters, training process, evaluation of metrics, and visualization of test images are implanted and prepared for using TF-GraF. Since this layer is very back-end of the server-side implementation, majority of this part demands programming knowledge. Therefore, this generally cannot be manipulated by users, but administrator can if there is a user request for customization or needs of maintenance.

*2.4. Client-side implementation*

Client-side is where users use the TF-GraF. This is the start point of the deep learning process use cases and end point at the same time. In this client-side, users can simply use TensorFlow object detection API without reading a single line of code by just clicking buttons on GUI (Figure 3). The TF-GraF are composed of six parts which are toolbar view (Figure 3, a; red box), display current directory and files view (Figure 3, b; orange box), control training related steps view (Figure 3, c; yellow box), control server with command line (Figure 3, d; green box), display selected image view (Figure 3, e; blue box), and display current directory path and activated environment view (Figure 3, f; purple box). Through this program, TF-GraF, user can upload, download, and delete data sets to server (Figure 3, a), the uploaded files in server-client can be controlled in TF-GraF (Figure 3, b). The TF-GraF can perform preprocessing such as creating of labelmap files (Figure 4), converting to tfrecord files based on annotation and image files (Figure 5), additionally, the TF-GraF can regulate model selection such as architectures and backbones (Figure 6, a), number of training steps (Figure 6, b), and hyperparameters (Figure 6, c). In addition, the TF-GraF contains evaluation functions such as various metrics analysis and visualization of test images based on trained models TF-GraF user manual was appended in github (https://github.com/boguss1225/ObjectDetectionGUI). In summary, the main function of client-side is triggering or transferring commands into the server-side and displaying result of server-side output. The user clicks buttons to trigger an assigned command so that the button can replace command line interface.

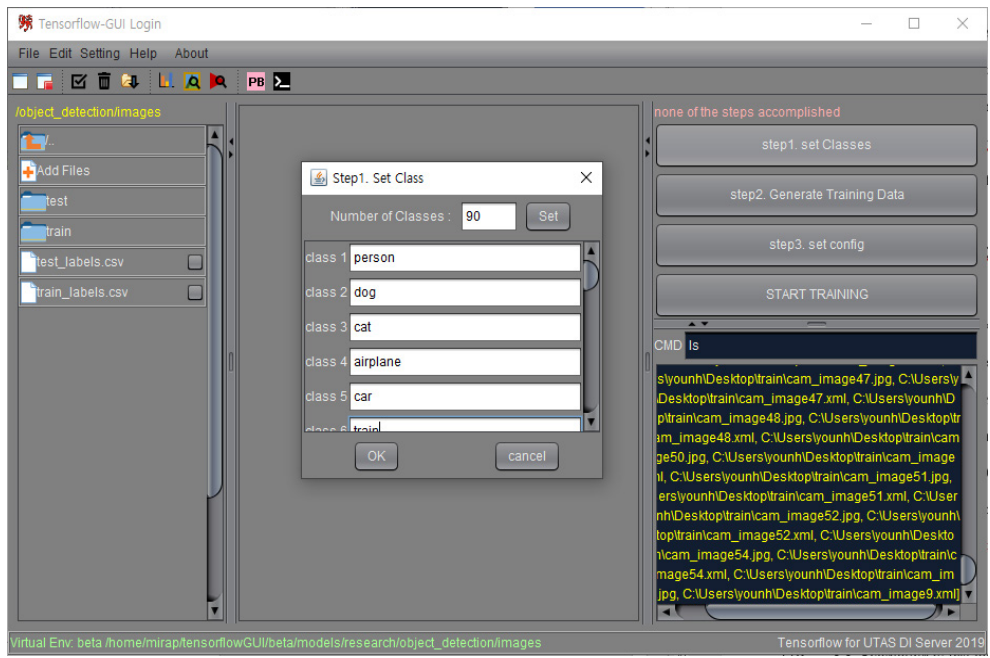

**Figure 4.** Creating of labelmap files in TensorFlow with user friendly Graphical Framework for object detection API (TF-GraF)

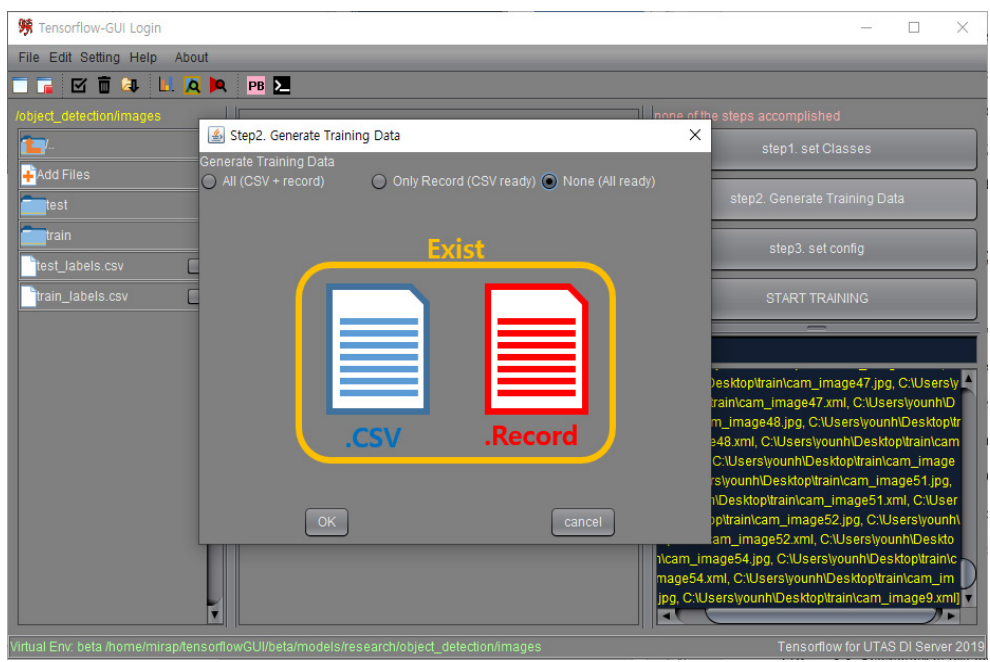

**Figure 5.** Converting tfrecord files from annotation files (both possible xml and csv format) and images in TensorFlow with user friendly Graphical Framework for object detection API (TF-GraF)

## 3. Preprocessing, training and evaluation of COCO dataset using TF-GraF

*3.1. Data preparation and preprocessing using TF-GraF*

   Images and annotations for training and testing were downloaded from common objects in context (COCO) dataset. The dataset uploading (Figure 3, a), creating of labelmap files (Figure 4), and converting to tfrecord format (Figure 5) were performed in TF-GraF. The data were divided with eighty percentage of train data. Fifty percentages train dataset were treated with random horizontal and flip, adjusting of brightness, contrast, and saturation, and 90 degrees rotation and all augmentation processes were performed using TF-GraF.

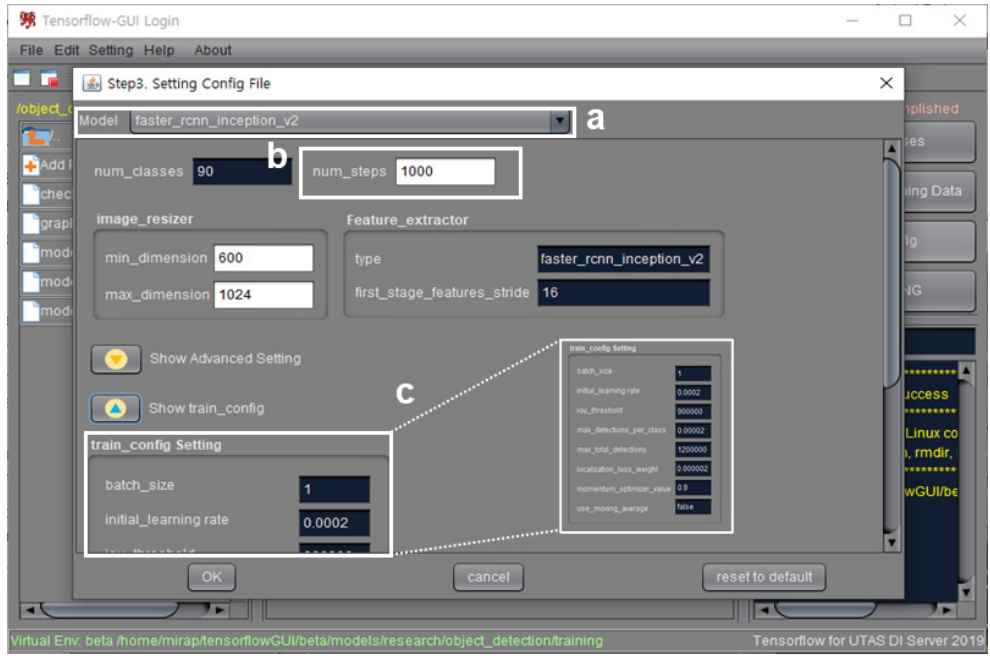

**Figure 6.** Control of architectures and backbones (a), step number (b), and various hyperparameters (c) for model training in TensorFlow with user friendly Graphical Framework for object detection API (TF-GraF).

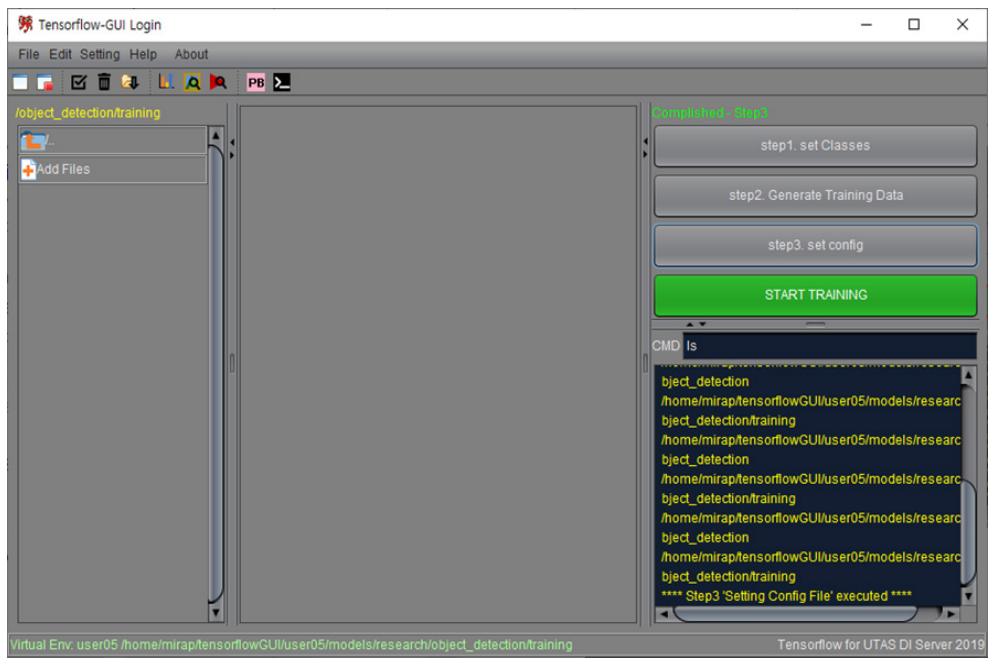

**Figure 7.** Model training, evaluation, and creation of inference files process in TensorFlow with user friendly Graphical Framework for object detection API (TF-GraF). Button of training (Green box)

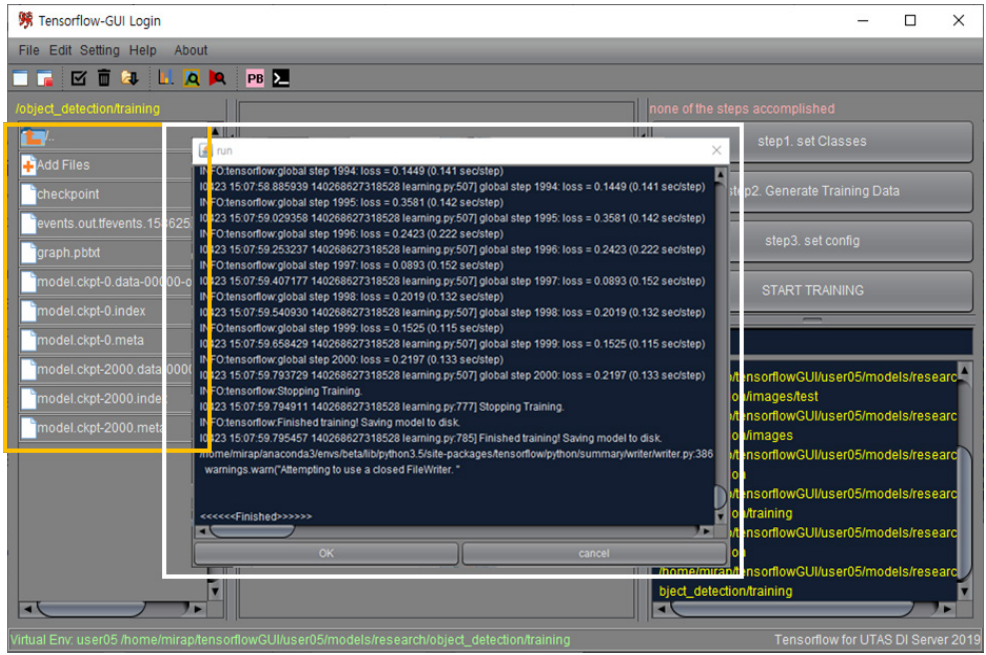

**Figure 8.** Creation of checkpoint files after training finished (yellow box) and detection process at real-time (white box) using TensorFlow with user friendly Graphical Framework for object detection API (TF-GraF).

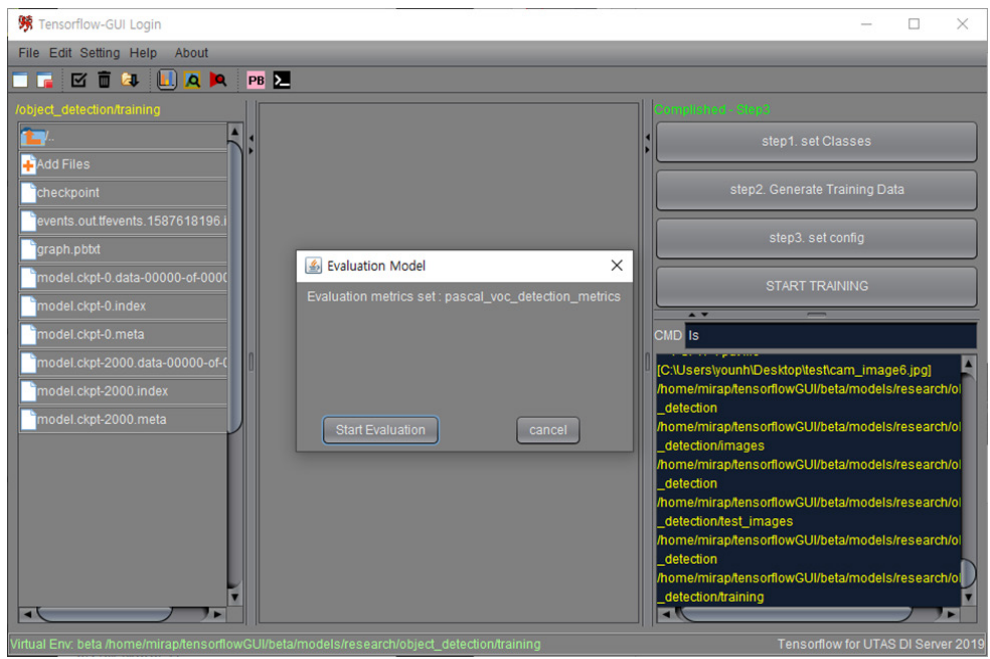

**Figure 9.** Performance evaluation by pascal VOC detection metrics for analysis performance (mean average precision; mAP) using checkpoint files in TensorFlow with user friendly Graphical Framework for object detection API (TF-GraF).

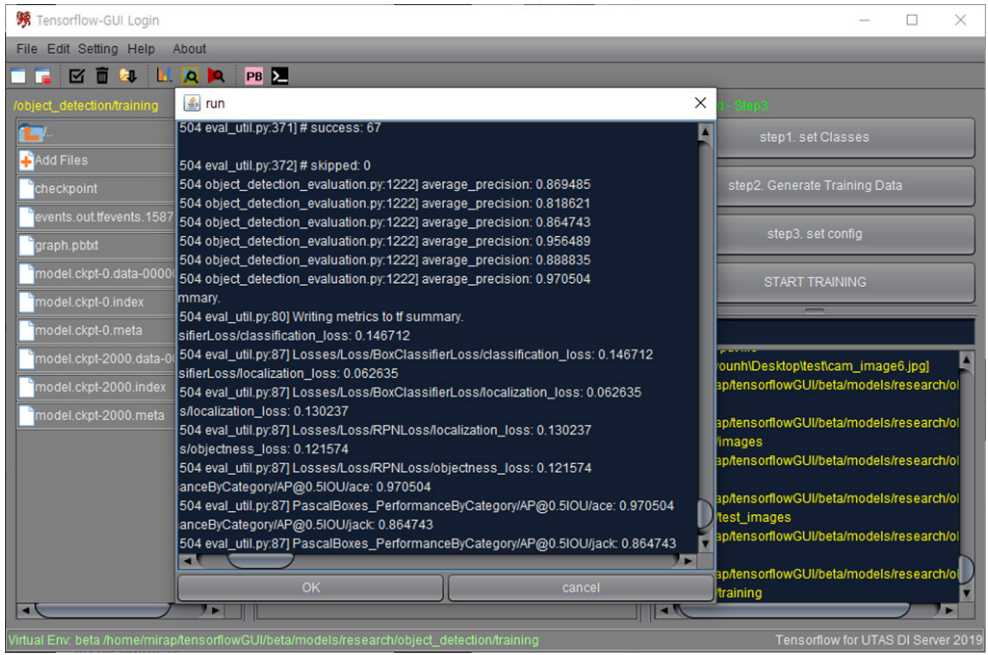

**Figure 10.** Evaluation result of performance in trained models by TensorFlow with user friendly Graphical Framework for object detection API (TF-GraF).

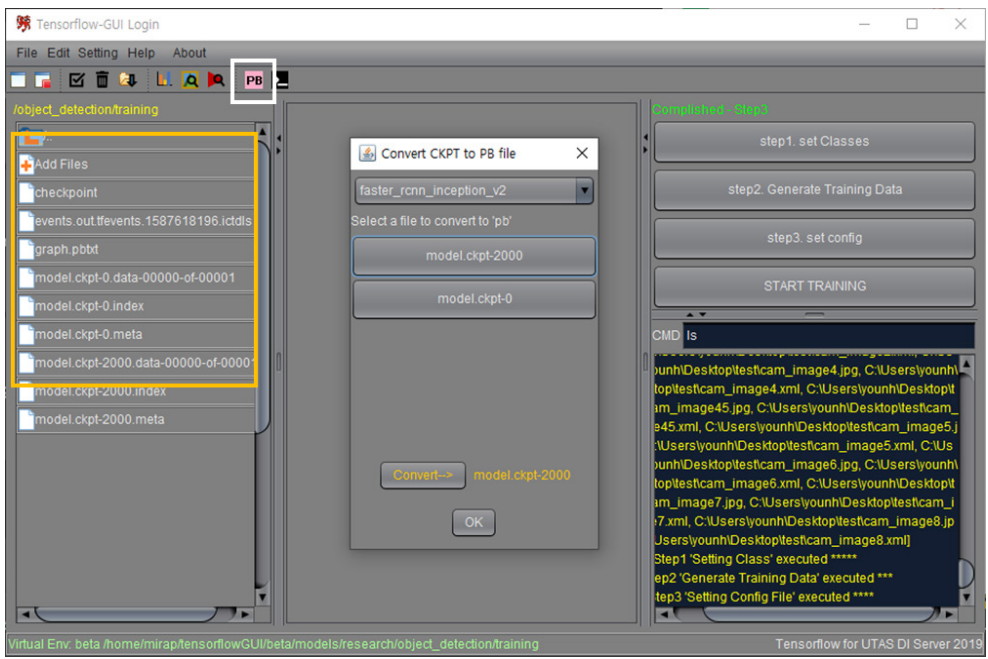

**Figure 11.** Converting checkpoint files (yellow box) to inference graph files in Pb converting function (white box) in TensorFlow with user friendly Graphical Framework for object detection API (TF-GraF).

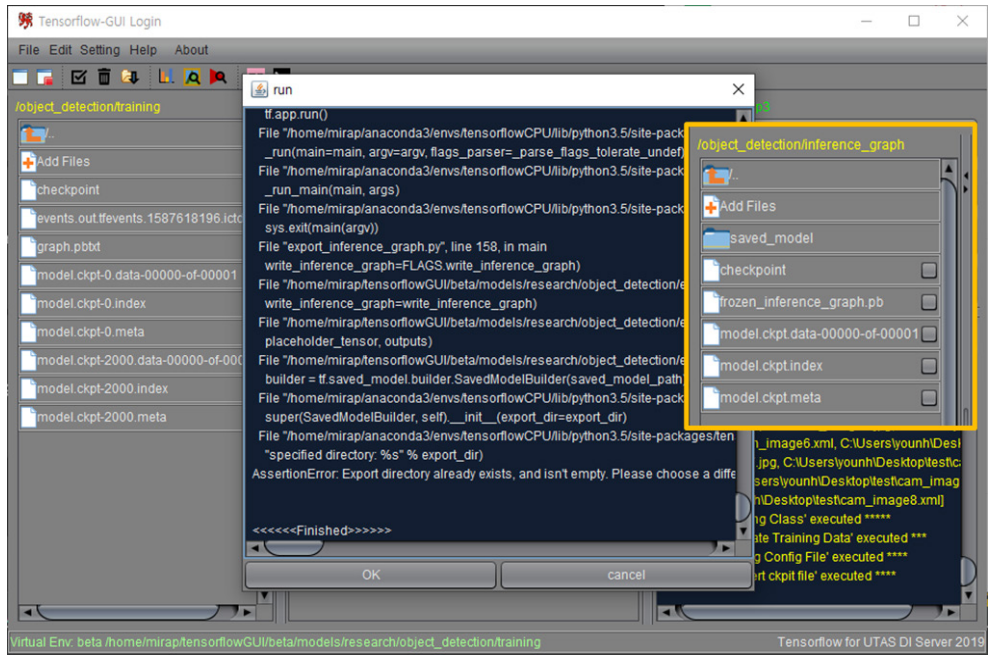

**Figure 12.** Created model inference graph files (yellow box) for image detection in TensorFlow with user friendly Graphical Framework for object detection API (TF-GraF).

*3.2. Training dataset using TF-GraF*

The data were trained using four architecture models which were SSD [19], Faster R-CNN [20], RFCN [21], and Mask-RCNN [22] and the backbone of SSD was used by MobileNet v2 [23], Faster R-CNN, RFCN, and Mask R-CNN were used by ResNet 101 [24]. Training were performed until 200,000 steps, and learning rate was 0.0002 when training was started and 0.00002 since 150,000 steps. All process of selection of architectures, backbones, setting of hyperparameter, and training configuration were performed in TF-GraF. TensorFlow 1.14 was implanted in python 3.5 Anaconda version in Ubuntu 16.04 LTC at server-side environment. The TF-GraF was installed at windows 10 in client-side environment. The training was initiated in TF-GraF (Figure 7, green box) and trained checkpoint files according to training steps (Figure 8, yellow box) and training processing (Figure 8, white box) were detected at real time in TF-GraF.

*3.3. Evaluation of test dataset using TF-GraF*

The checkpoint files by SSD MobileNet v2, Faster R-CNN ResNet 101, RFCN ResNet 101, and Mask R-CNN ResNet 101 were evaluated mean average precision (mAP) using COCO metrics in TF-GraF function (Figure 9). After, we detected the results of metrics in separated window in TF-GraF (Figure 10). In addition, to visualize test images by trained models, the checkpoint files (Figure 11, yellow box) were created pb files (Figure 12, yellow box), using converting of inference graph function (Figure 11, white box) in TF-GraF. After then, detected boxes and segmentation in test images (Figure 13A) were visualized using pb files (Figure 12, yellow box) based on trained models such as SSD MobileNet v2 (Figure 13B), Faster R-CNN ResNet 101 (Figure 13C), RFCN ResNet 101 (Figure 13D), and Mask R-CNN ResNet 101 (Figure 13E) in TF-GraF.

**4. Discussion and further study**

*4.1. Need external annotation tool for labeling*

To be able to use the proposed system, it is essential to have annotated data set which consist of set of images and xml files. To generate the data set, however, user need to use external annotation tool for annotation. Even though there are various available annotation tools on internet, those tools will require other dependencies to be able to utilize TF-GraF. This can increase complexity of TF-GraF usage and decrease usability for non-IT background users.

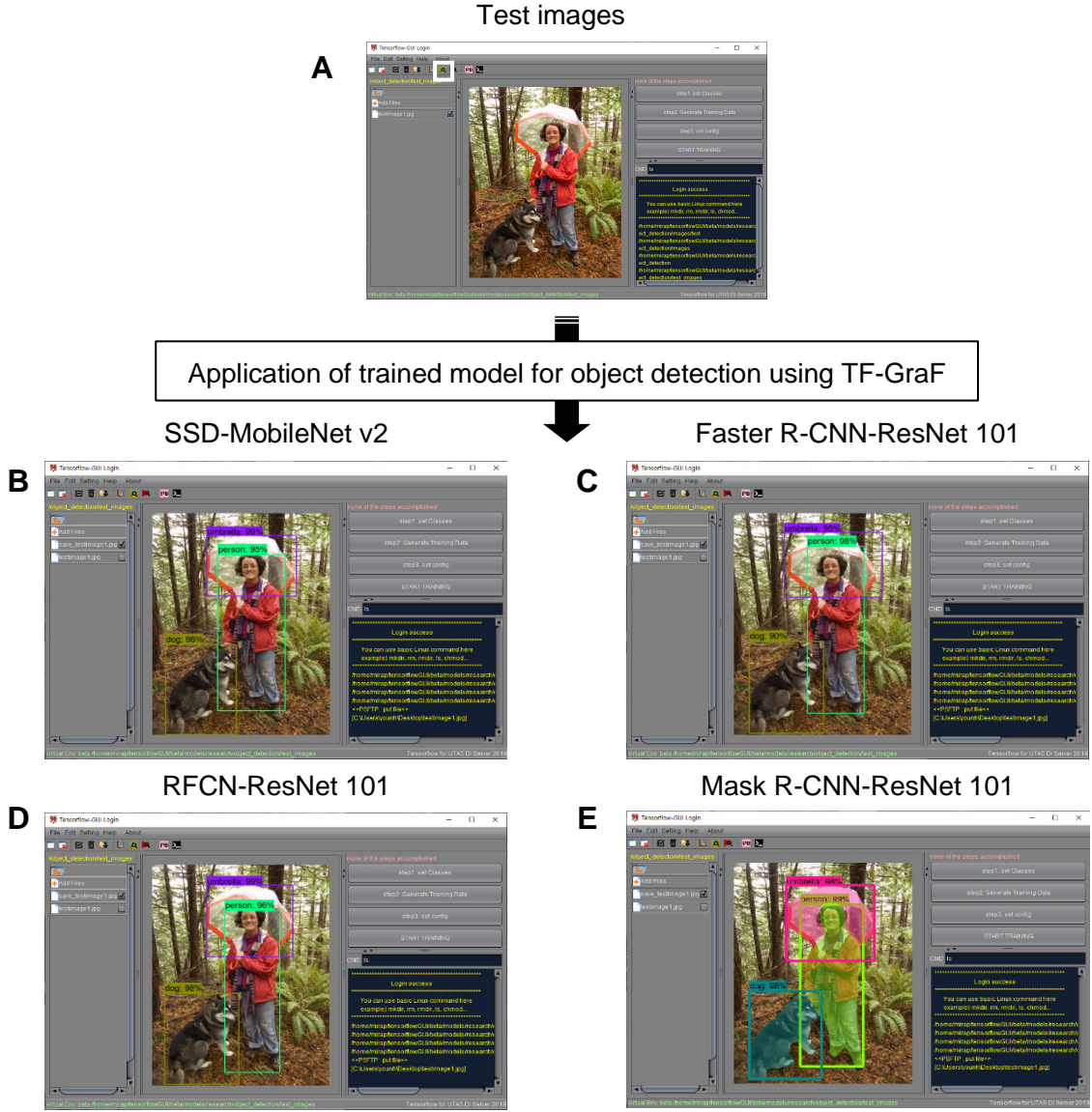

**Figure 13.** Object visualization of test images using trained models in TensorFlow with user friendly GRAphical Framework for object detection API (TF-GraF). Preparation of test image and click the object detection function (A, white box), detected boxes based on deep learning trained models of SSD-mobileNetv2 (B), Faster R-CNN-ResNet 101 (C), and RFCN-ResNet 101 (D). Detected boxes and segmentation based on Mask R-CNN-ResNet 101 (E)

*4.2. Hyperparameter setting*

There are various functions and configurable options in TensorFlow object detection API. However, user only allowed to use functions that is accessible within GUI use cases. This limit further functionality and configuration of TensorFlow usage. To overcome this limitation, user can request customizing their environment to administrator or upgrading GUI program to developer. These requests, however, still need certain amount of time to be accomplished.

*4.3. Resource collision between users*

In the suggested system, users share GPU resources to train models with other users in the same server. In that nature, it is very common to get collision between users since the GPU resources are limited. Generally, the collision happens when a user tries to obtain GPU resource which is already occupied by some other user.

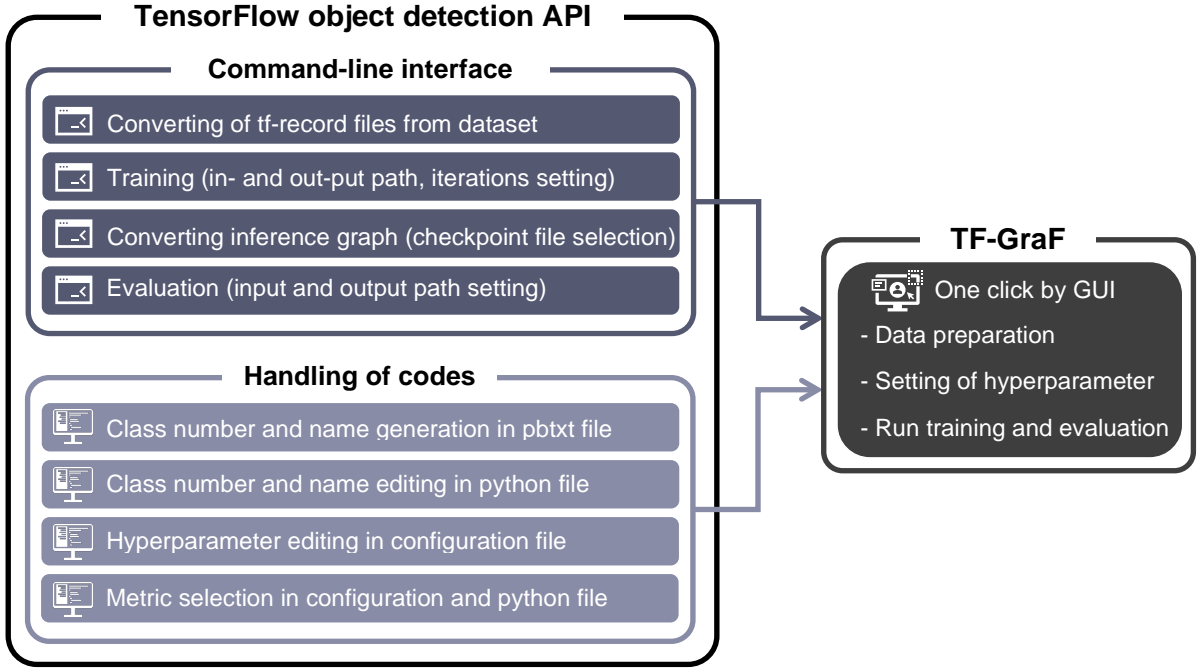

**Figure 14.** Overview of TensorFlow with user friendly Graphical Framework for object detection API (TF-GraF).

If that is the case, it seems to be able to overcome the limitation by implementing extra resource scheduling system or just simply adding GPU resources. This can be same issue on CPU resources since CPU is also limited resource in server side. During the implementation test, however, CPU does not seem to be a problem for collision because those process consuming CPU power is relatively simple and short. But training sessions are relatively complex and time consuming. In consequence, those processes which need GPU resource have high possibility to get collision with another user if there are the number of users in the system.

*4.4. Data Centralization*

Since The proposed system is operated in server side, user must upload data to the server. This cause inevitable consequence of data centralization. If user want to train with something unsensitive data from ethical or security issue, data centralization would not be issue. However, if there is a user who want to train images such as sets of private medical image or security sensitive data, data centralization can be major issue to give up using the system.

*4.5. Implementation of high level functions*

TensorFlow object detection API is supporting high level performances beside mentioned functions of this paper, of these, building of custom CNN models and augmentation are representative functions. The functions are independently designed using python files of core folder in TensorFlow object detection API which are flexibly integrated with existing code files. For example, if user need to use state of the art CNN model or apply custom augmentation such as complex of some degree rotation, gaussian filter, and brightness, they should make new files in core folder in TensorFlow object detection API. However, this whole process is difficult for general user, we are going to add the functions in the future.

**5. Conclusion**

The TF-GraF allows anyone, even without any previous knowledge of deep learning frameworks, to design, train and deploy machine intelligence models without coding (Figure 14). It simplifies and accelerates the process of working with the deep learning so that it can be more productive and efficient process in terms of saving time. Since TF-GraF takes care of setting and configuration, it allows anyone to contribute to their project without spending time to install complex software and environment. It unifies all workflows with simplified GUI and provides good performance for building, deploying, and testing procedures by a few clicks.

Therefore, usage of TF-GraF, many researchers who do not have understanding of computer science and IT can work with the fastest way to train neural nets for object detection based on the deep learning without programming.

**Data availability**
Implementation of TensorFlow object detection API for development of TF-GraF
https://github.com/TensorFlow/models/tree/master/research/object_detection.
COCO dataset for train and test
https://cocodataset.org

**Code availability**
The code of TF-GraF for TensorFlow object detection API is opened at
https://github.com/boguss1225/ObjectDetectionGUI

**Supplemental data**
TF-GraF user manual (https://github.com/boguss1225/ObjectDetectionGUI)